\documentclass{aa}    
\input psfig.tex
\begin{document}
\title {S~111 and the polarization of the B[e] supergiants in the Magellanic 
Clouds\thanks{Based on observations obtained at the Cerro Tololo 
Inter-American Observatory}}

\titlerunning {B[e] supergiants in the Magellanic Clouds}
\authorrunning {Melgarejo et al.}

\author{Roc\'{\i}o Melgarejo\inst{1}, 
A. M\'ario Magalh\~aes\inst{1}\fnmsep\thanks{Visiting Astronomer, Cerro Tololo Inter-American Observatory. CTIO is operated by AURA, Inc. under contract to the National Science Foundation},
Alex C. Carciofi\inst{1} 
\and Cl\'audia V. Rodrigues\inst{2}}

\offprints{R. Melgarejo}

\institute{Dept. de Astronomia, Instituto Astron\^omico e Geof\'{\i}sico,
Universidade de S\~ao Paulo, Caixa Postal 3386, 01060-970 S\~ao Paulo, SP, Brazil\\
\email{rocio@iagusp.usp.br, mario@iagusp.usp.br, carciofi@usp.br}
\and Div. Astrof\'{\i}sica, Instituto Nacional de Pesquisas Espaciais/MCT, Caixa
Postal 515, 12201-970, S\~ao Jos\'e dos Campos, SP, Brazil\\
\email{claudia@das.inpe.br}}

\date{Received 29 May 2001; accepted 7 August 2001}

\abstract{
We have obtained linear polarization measurements of the Large Magellanic 
Cloud B[e] supergiant S~111 using optical imaging polarimetry. The intrinsic 
polarization found is consistent with the presence of an
axisymmetric circumstellar envelope. We have additionally estimated the electron
density for S~111 using data from the literature and
revisited the correlation between polarization and envelope parameters
of the B[e] supergiant stars using more recent IR calibration color data. The
data suggest that the polarization can be indeed explained by electron scattering.
We have used Monte Carlo codes to model the continuum
polarization of the Magellanic B[e] supergiants. The results indicate that the
electron density distribution in their envelopes is closer to a homogeneous
distribution rather than an r$^{-2}$ dependence. At the same time, 
the data are best fitted by a spherical distribution with density contrast
than a cylindrical distribution. The data and the model results
support the idea of the presence of an equatorial disk and of the two-component
wind model for the envelopes of the B[e] supergiants.
Spectropolarimetry would help further our knowledge of these envelopes. 
}

\maketitle
%

\section{Introduction}

In the HR diagram, there is a temperature dependent upper limit for the stellar 
luminosity (Humphreys \& Davidson 1979). Among the stars located around that 
region, we find the Luminous Blue Variables (LBVs), Wolf-Rayet stars and the
 less well understood B[e] supergiants
(B[e]SG). B[e]SG are stars with permitted and forbidden lines in their optical
spectra dominated by strong Balmer emission lines, sometimes with P Cygni
profiles. A characteristic of this group is a strong infrared excess
that provides evidence for hot circumstellar dust. These stars show
simultaneously narrow emission lines in the optical and
broad absorption lines in the UV. Such hybrid spectra can be
explained by a two-component wind model (Zickgraf et al. \cite{z85}, hereafter
Z85). This wind consists of a hot, fast, polar wind and a cool, slow, dense
wind in the equatorial region. Recent reviews about B[e] stars can be found in 
Lamers et al. (1998) and Zickgraf (1999).

At present, fifteen stars of this class are known in the MC and eight in the
Galaxy. Data on the MC B[e]SG are still badly needed. For instance, B[e]SG do 
not show in general photometric
variability ($\Delta$m $<$ 0.1 mag). However, R~4 (a B[e]SG in the SMC)
exhibited
a typical LBV variation in the optical (up to 0.7 mag, Zickgraf et al. 1996a).
In addition, the luminosities of the four new B[e]SG, discovered by Gummersbach 
et al. (1995) in the LMC, extend down to log $L/L_{\sun} = 4$, lower than those 
of the already known B[e]SG.

Magalh\~aes (\cite{m92}, hereafter M92) obtained polarization
measurements of nine B[e]SG in the Magellanic Clouds (MC), and showed evidence 
for the presence of non-spherically symmetric envelopes, giving support for the
two-component wind model. The available data correlated some with the average 
electron density of the envelope, $N_{e}$, but correlated slightly better with the
infrared $[K-L]$ dust excess. M92 described ways in which polarimetry of B[e]SG 
could be useful in probing their envelopes.

HDE 269599s (= S~111 = Sk~147a, Henize 1956 and Sanduleak 1968, respectively) 
is a B[e]SG in the LMC that belongs to a compact
cluster ($\sim$10$\arcsec$) with about 10 stars (Appenzeller et al. 1984).
This star could not be observed with photoelectric polarimetry by M92. It has
however later been observed by us
with CCD imaging polarimetry. In this paper, we add S~111 to the group
of MC B[e]SG with known polarimetry (Sect.~2). After estimating the envelope
electron density of S~111
we revisit the correlations between B[e]SG polarization and $N_{e}$
and dust excess in the light of newer observational data in the literature
(Sect.~3). We then model the
intrinsic polarization of these stars by using Monte Carlo codes and try to
infer properties of the B[e] envelopes (Sect.~4). Conclusions are presented in the last
section.


\section{Observations and data reduction}

We obtained optical linear polarization images of S~111 at
the 1.5 m telescope of CTIO in December 1991. We used the CCD Tektronix
1024x1024 Tek\#1 camera and the measurements were
made through a $V$ filter. The polarimeter is a
modification of that direct CCD camera to allow for high
precision imaging polarimetry. It consists basically of a rotateable half-wave
plate followed by a fixed analyser and a filter. The arrangement is described
in more detail by Magalh\~aes et al. (\cite{m96}).

As analyser we used a custom-made,
44 mm wide square double calcite block
(built at Optoeletr\^onica, S\~ao Paulo). Each component prism was
cut with its optical axis at $45\degr$ to their faces and
they were cemented with
their optical axes crossed. This arrangement minimizes the astigmatism and
color
which are present when a single calcite block is used
(Serkowski \cite{ser74}).
This Savart plate gives us two images of each object in the field, separated by
1 mm (corresponding to about 18\arcsec\ at the telescope focal
plane) and with orthogonal polarizations.
By rotating the waveplate between CCD exposures, the ratio of the flux in each
of the object's images changes with an amplitude which is proportional to
the incoming beam's polarization. One polarization modulation cycle is covered 
for every $90\degr$ rotation of the waveplate. The
simultaneous observations of the two images allows observing under
non-photometric conditions at the same time that the sky polarization is
practically cancelled (Magalh\~aes et al. \cite{m96}).

CCD exposures were taken with the waveplate
rotated through 8 positions $22\fdg5$ apart.
The exposure time at each position was 60 s.
After bias and flatfield corrections,
photometry was performed on the images of objects in the field with IRAF,
and then a special purpose FORTRAN routine processed these data files
and calculated the normalized linear polarization
from a least squares solution. This yields the
Stokes parameters $Q$ and $U$ as well as the theoretical (i.e., photon noise)
and measurement errors. The latter are obtained from the residuals of the
observations at each waveplate position angle (${{\psi}_i}$) with regards to the
expected cos $4{{\psi}_i}$ curve and
are quoted in Table 1; they are consistent with
the photon noise errors (Magalh\~aes et al. \cite{m84}).
The instrumental $Q$ and $U$ values were
converted to the equatorial system from standard star data obtained in the same
night. The instrumental polarization was measured to be less than 0.03\%.

To obtain the intrinsic polarization we need to know the foreground
interstellar polarization. In order to do this, we employed field stars. The
sample that we used as field stars corresponds to three stars in the
same cluster as S~111 (Fig. 1). The advantage of using them is because
we know that
those stars are at the same distance from us as S~111 as well as angularly
close to it. The intrinsic polarization is calculated as follows:
\begin{eqnarray}
Q_{\rm{i}} & = & Q_{\rm{o}} - Q_{\rm{IS}}, \nonumber \\ & & \\
U_{\rm{i}} & = & U_{\rm{o}} - U_{\rm{IS}} \nonumber
\end{eqnarray}
where the $(Q_{\rm{IS}},U_{\rm{IS}})$ values are weighted mean of the three
stars. Subscripts i, o and IS correspond to intrinsic, observed and interstellar 
polarization, respectively.

In Table 1 we show the observed polarization measurements 
($P = \sqrt{Q^{2}+ U^{2}}$ ) for four stars of the
cluster. The stars labeled with 2, 3, 4 in Fig. 1 were used as
field stars to calculate the foreground interstellar polarization. S~111
is star 1. In Table 2 we present the optical polarization data
for S~111. The last 3 columns of Table 2 show that S~111 has intrinsic optical 
polarization.

\begin{figure}
\psfig{figure=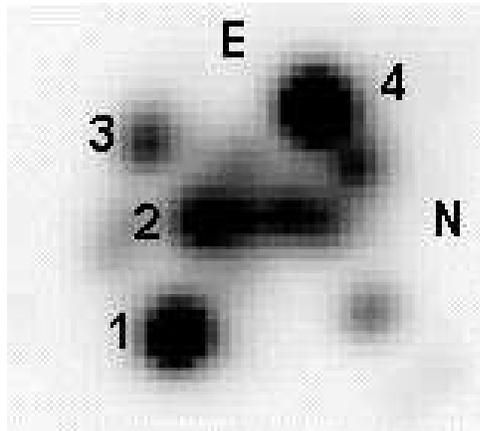,width=6.5cm}
\caption{The HDE 269995 cluster in the LMC, from a section of an image taken in 
CTIO. S~111 is star number 1. The others were used for determining the foreground 
interstellar polarization towards S~111.}
\label{}
\end{figure}

\begin{table}
\caption{Observed Optical Polarization Data for the Cluster}
\medskip
\noindent
\begin{tabular}{cccc}
\hline
\noalign{\smallskip}
N & $P$ (\%) & $\sigma_P$ (\%) & $\theta$ ($\degr$)  \\
\noalign{\smallskip}
\hline
\noalign{\smallskip}
1 (=S~111)  &  0.710  &  0.033  &  50.5   \\
2  &  0.672  &  0.085  &  26.1   \\
3  &  0.532  &  0.075  &  35.8   \\
4  &  0.544  &  0.029  &  29.7   \\
\noalign{\smallskip}
\hline
\end{tabular}
\renewcommand{\arraystretch}{1}
\end{table}

\begin{table*}
\caption{Optical Polarization Data for S~111}
\medskip
\noindent
\begin{tabular}{ccccccccccc}
\hline
\noalign{\medskip}
\multicolumn{3}{c}{OBSERVED} & &
\multicolumn{3}{c}{FOREGROUND} & &
\multicolumn{3}{c}{INTRINSIC} \\
\noalign{\smallskip}
\cline{1-3} \cline{5-7} \cline{9-11} \\
$P$ (\%) & $\sigma_P$ (\%) & $\theta$ ($\degr$) & &
$P$ (\%) & $\sigma_P$ (\%) & $\theta$ ($\degr$) & &
$P$ (\%) & $\sigma_P$ (\%) & $\theta$ ($\degr$) \\
\noalign{\smallskip}
\hline
\noalign{\smallskip}
0.710 & 0.033 & 50.5 & &
0.552 & 0.026 & 30.0 & &
0.467 & 0.042 & 75.3 \\
\noalign{\smallskip}
\hline
\end{tabular}
\renewcommand{\arraystretch}{1}
\end{table*}


\section{Polarization and the B[e]SG envelopes}

The envelopes of B[e]SG provide an ample opportunity for scattering of radiation
from the central star due to the available free electrons, as revealed from
their emission lines and near IR ($J$ band) excess, and dust, evidenced by their 
IR excess. Following M92, we reinvestigate the correlations between the optical
polarization of B[e]SG and inferred electron densities and infrared excesses.
This reanalysis is warranted by the additional data point S~111 provides,
the recent finding concerning the binarity of AV~16/R~4 and newer photometric
calibration tables for optical-near IR photometry.

\subsection{Polarization and Electron density}

It is believed that the $J$ excess from B[e]SG is due to free-free and
bound-free radiation. Therefore, such excess is a measure of the electron
density in the envelopes of B[e]SG (Zickgraf et al.
\cite{z86}, hereafter Z86). M92 correlated the polarization of nine MC B[e]SG with 
the electron density ($N_e$) obtained from $J$ excesses ($\Delta J$) by Z86.
A measure of the optical magnitude needed to fit the stellar continuum energy 
distribution, and hence for obtaining $\Delta J$, is not available for S~111. Thus we 
estimated $N_e$ for S~111 not from $\Delta J$ but from the emission measure
(EM) of H$\alpha$. 

For estimating $N_e$ from H$\alpha$ we followed the method outlined by Z86. 
As a check on our procedure, we initially applied it to the other B[e]SGs, for 
which optical
magnitudes are available. The first step required is obtaining the H$\alpha$ 
continuum flux. This was done
using an interpolation of the fluxes in the $V$ and $I$ filters using
the observed magnitudes of Z85, Z86 and Zickgraf et al. \cite{z89} (hereafter
Z89)
and a power law, $F_\lambda= k\lambda^{-\rm{n}}$. The exponent n was found to 
vary between 1.5 and 2.9 (with median = 2.5) for the different stars. The magnitudes
were corrected for interstellar extinction, using the extinction curve of Code 
et al. (1976). The Galactic color excess $E(B-V)$ was taken as 0.05 (Z85, 
Bessell 1991 and Rodrigues et al. \cite{r97}) and the Magellanic 
color excess $E(B-V)$ from Z86 and Z89. The
fluxes were then computed using the absolute fluxes of Bessell (1979).
Our results compared well to those of Z89 within 30\%.

For S~111, in order to calculate the H$\alpha$ continuum flux we estimated the
magnitude of the object in the $V$ filter, using the bolometric
correction $BC_V$ and the stellar luminosity by
McGregor et al. (\cite{m88}, hereafter M88), as well as the distance modulus for the
LMC (Walker 1998). The estimated intrinsic (i.e., extinction corrected) 
apparent magnitude for S~111 thus found was
9.20 mag. This estimated $V$ magnitude is not quite consistent with the near
infrared magnitudes of M88 (their Table 5). We did
perform
another estimate of the magnitude of S~111 using the photometry of Mendoza 
(1970), who obtained the magnitude in the $V$ filter for the cluster to
which S~111 belongs. Our CCD images provide the
relative fluxes of the brightest stars in the cluster. It was then possible to
estimate the magnitude of S~111 from the cluster $V$ magnitude of 
Mendoza (1970) and our relative flux data. For this purpose, we assumed that
the flux of the cluster was due to the 
contribution of its four brightest stars. From this
assumption we obtained a new intrinsic magnitude for S~111, 10.26 (with an 
error of $\sim$0.02, the error in V for the S~111 cluster as estimated by Mendoza), 
a value more consistent with the near infrared magnitudes of S~111.
Since there is no available $I$ filter measurement
for S~111 either, the H$\alpha$ continuum flux of S~111 was obtained from
an extrapolation of this $V$ flux assuming a value of n=2.5 for the index of the continuum
flux distribution. A more definite estimate should probably be best done from direct filter 
photometry for S~111, specially in the $I$ filter.

The next step was obtaining the H$\alpha$ flux (H$\alpha$ continuum flux
$\times$ equivalent width) and then the H$\alpha$ luminosity. For S~111, the
equivalent width was measured from the H$\alpha$ profile given by Stahl \& Wolf
(1986) as 356~\AA. The EM was calculated using the case B recombination theory formula
(Osterbrock 1989). From the definition of EM, $N_e$ can be calculated as
$\sqrt{\rm{EM}/\rm{V}}$,
where V is the volume of the envelope. The shape of the envelope was taken
as suggested by Z85. It consists of a disk with inner radius
of 1~$R_{\star}$, outer radius of 300~$R_{\star}$ and thickness of 1
$R_{\star}$. For the stellar radius of S~111, we obtained the
value of 148~$R_{\sun}$ using the effective temperature and stellar luminosity
from M88. The resultant $N_e$ for S~111 was 0.43~$\times$~$10^9$~cm$^{-3}$.

As a check on the procedure, we reevaluated $N_e$ for other B[e]SG.
The H$\alpha$ equivalent width of other B[e]SG stars were
either taken from the literature (R 126, S 134 and S 12; Z89) or estimated by us
(S 18, S 22 and R 66). Among the latter, two compared very well with the 
literature (S 18, Z89; S 22, Schulte-Ladbeck and Clayton 1993).
Both the fluxes and EM values 
obtained by us agreed with those of Z86 and Z89 estimated from H$\alpha$.
On the other hand, our values of $N_e$ differ some, though always of the same 
order of magnitude, from those of Z89's obtained from $J$ excess. In general, 
the electron density from $\Delta J$ is on average two times larger than that of 
H$\alpha$. We hence used 0.8 $\times$ $10^9$ cm$^{-3}$ for the electron density 
in the envelope of S~111.

As discussed in detail by M92, high levels of intrinsic
polarization are related to edge-on stars. We hence expect that polarization
of these stars might be proportional to the electron density and the other stars
(pole-on and intermediate-on) be distributed in a triangular pattern due to the
dependence on $\sin^2$i, where i is the inclination angle through which the discs are
seen. In Fig. 2 we plot the intrinsic polarization as
a function of electron density. We use our $N_e$ value above for S~111 and Z89's for 
the remaining stars. The polarization values are taken from Sect. 2 of 
this paper for S~111 and from M92 for the remaining objects.

Unlike M92, we have omitted from the plot in Fig. 2 the SMC star AV~16/R~4. 
Zickgraf et al. (1996a) have 
shown that this object is in fact a binary star, consisting of a B[e] star and 
an early A-type companion. The observed polarization of an object is basically 
the ratio between the scattered, polarized light from the
circunstellar material and the total light from the system. Hence, such source
of additional, unpolarized light in the system (i.e., the A component) would act 
to dilute an otherwise larger intrinsic polarization. In fact, of the four 
equator-on cases (as evidenced by their spectroscopic 
properties from the work of Zickgraf and collaborators) analysed by M92, R~4 
showed the smallest intrinsic polarization. Interestingly, in addition to that, in the $P$ vs. 
$N_e$ correlation R~4 of M92 (his Fig. 4) R~4 clearly stood out from the other 
three equator-on cases by being well off the line defined by those objects.

We did a linear $\chi^2$ fit for the remaining spectroscopically edge-on stars 
(R~50, R~82 and S~22, filled squares in Fig. 2) and present that fit in Fig. 2. 
It can be seen that the remaining stars, i.e., the pole-on and intermediate stars, 
are distributed below this fit line, as 
expected. As in M92, we chose to use the Spearman rank-order correlation 
coefficient ($r_{s}$), to which a statistical significance can be attached. 
The calculated value of the $r_{s}$ for the {\it P} 
versus {\it N$_{e}$} correlation for all objects is 0.55, with a 12\% 
probability, where a small probability means high significance. 
This indicates a moderate correlation between $P$ and $N_{e}$. 
Nevertheless this correlation is much higher than that of M92 (0.32 with 41\%),
which included R~4.

\begin{figure}
\psfig{figure=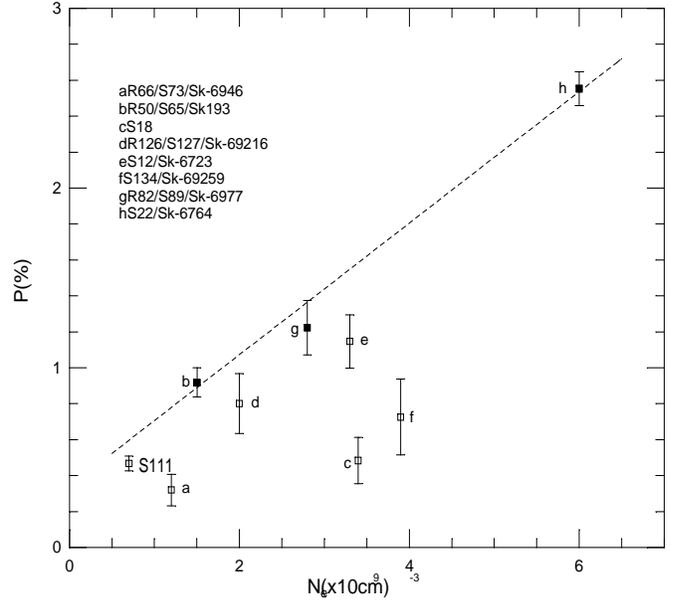,width=9cm,height=8cm} 
\caption{Intrinsic polarization versus electron density. The dotted line 
corresponds to a $\chi^2$ linear fit for edge-on stars (R~50, R~82 and S~22), 
indicated by the filled squares.}
\label{}
\end{figure}

\subsection{Color excess $\Delta(K-L)$}

We can use the near-infrared color excess at wavelenghts longer than $J$ as a 
measure of dust emission (Z86, Z89). We define the following color excess $\Delta(K-L)$:

\begin{equation}
\Delta(K-L) = (K-L)_{\rm{corr}} - (K-L)_{\rm{int}} \,,
\end{equation}

\noindent where $(K-L)_{\rm{corr}}$ is the observed color index $(K-L)$
corrected for interstellar extinction and $(K-L)_{\rm{int}}$ is the intrinsic color index
due to the stellar contribution.

$(K-L)_{\rm{int}}$ was taken from the newly available 
$UBVRIJHKL$ color tables of Bessell et al. (1998). These tables take into
account the star's gravity, finer intervals of effective temperature 
and colors for T $>$ 21000K compared to those of Johnson (1966). $(K-L)_{\rm{corr}}$ 
was obtained using the observed magnitudes of Z85, Z86 and
Z89, corrected by the interstellar color excesses taken from Z85 for the Galaxy and from Z86 
and Z89 for the MC and using the extinction coefficients of
Leitherer \& Wolf (1984). We note that there might
be probably uncertainties in the color excesses derived in these papers. The 
reason is that the dust envelope may also contribute to the observed 
reddening of the stellar spectrum. The amount of such reddening will 
depend on detailed modeling. The derived interstellar color excesses may then be 
considered as upper limits. One such example may be S 18, which has a derived 
$E(B-V)=0.40$ (Z89). Such a relatively large color excess, albeit possible, is rarely observed 
towards the SMC. Nevertheless, as expected, these color excesses have 
only a small effect in $(K-L)_{\rm{corr}}$ at these long, IR wavelengths.

In Fig. 3 we plot the intrinsic polarization as a function of the color excess
$\Delta(K-L)$. As before, we omit R~4 from the correlation. A linear 
$\chi^2$ fit to the spectroscopically edge-on
stars (R~50, R~82 and S~22) is also shown in the figure. It can be seen 
that not all stars are distributed below the fit
line. The Spearman correlation coefficient for the {\it P} versus 
{$\Delta(K-L)$} relation for all objects is 0.20, with a 60\% probability. This value of 
$r_{s}$ shows a very weak correlation between $P$ and $\Delta(K-L)$.

\begin{figure}
\psfig{figure=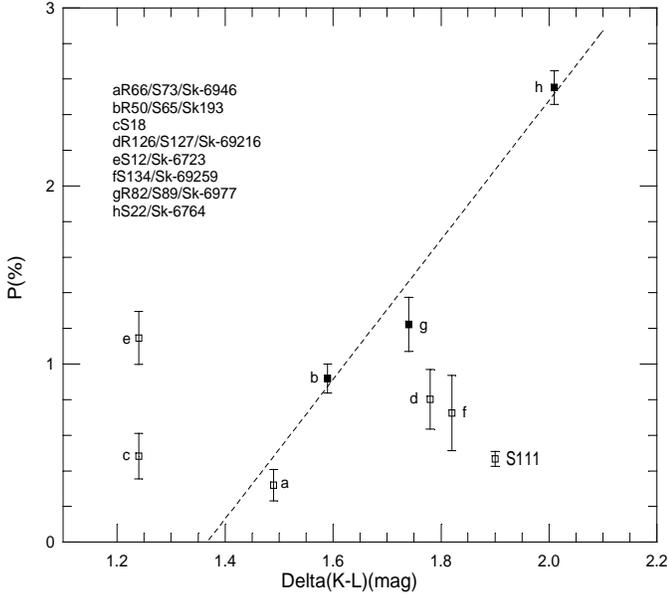,width=9cm,height=8cm} 
\caption{Intrinsic polarization versus color excess. The dotted line as in Fig. 2.}
\label{}
\end{figure}

In summary, the existing B[e]SG polarimetric data show that intrinsic polarization 
correlates better with electron density than dust scattering. This result is
consistent with the spectropolarimetric observations of S~22 by Schulte-Ladbeck 
\& Clayton (1993). They found that the polarizing mechanism in that object can be 
explained by electron scattering. Additional multifilter and 
spectropolarimetric measurements of the B[e]SG are clearly of the greatest 
interest. As discussed in M92, these measurements will allow an improved separation 
between intrinsic and foreground polarization as well as the detection of changes in 
polarization across spectral features produced in the B[e]SG envelopes. Such changes 
will help reveal the physical structure of their envelopes.


\section{Monte Carlo models}

In order to quantitatively infer consequences of the observed polarization
to the B[e]SG envelopes, we used a Monte Carlo code developed by our group 
(Carciofi \& Magalh\~aes 2001). In this preliminary study, two geometries
were chosen for these envelopes (although others may be used): cylindrical 
geometry and spherical geometry with density contrast. 
The density laws used were one of constant density and one proportional to $r^{-2}$. 
The codes perform the radiative transfer of the Stokes parameters ($I,Q,U,V$) 
for photons in scattering envelopes. 

From the preceeding section, we found that polarization correlates best
with electron density. We have hence only considered models with electron
scattering at this point. We used data on nine B[e]SG for the models; 
that is, those from M92, with the exception of the binary AV~16/R~4, 
plus S~111. In Table 3 we summarize their stellar radius (Zickgraf 1990) and electron 
density values (Z89). For the modelling we considered five values for the 
average electron density in the envelope which encompass 
the observed ones (Table~3): [0.5, 1.5, 3, 4, 6] $\times$ $10^9$ cm$^{-3}$. 
The mean electron density for each case is calculated from $<N_{e}^{2}>$ (=~EM/V), 
as described in Sect. 3.1. For the models, we have used the same approach, i.e., we
calculated $<N_{e}^{2}>$ for a given geometry and density law and then took the 
square root of this value to represent the mean electron density of that specific model. 
In the figures of the models we use ${<N_{e}^{2}> }^{1/2}$ along the x-axis.

\begin{table}
\caption{Electron Densities and Stellar Radii for MC B[e]SG}
\medskip
\noindent
\begin{tabular}{lcc}
\hline
\noalign{\smallskip}
Star & $R^{\rm a}$ $(R_{\sun})$ & $N_e$$^{\rm b}$ ($\times$ $10^9$ cm$^{-3}$) \\
\noalign{\smallskip}
\hline
\noalign{\smallskip}
S~111    &    148$^{\rm c}$  &   0.8$^{\rm c}$  \\
R~126    &     72    &   2.0    \\
S~134    &     45    &   3.9    \\
S~12     &     30    &   3.3    \\
R~66     &    125    &   1.2    \\
R~82$^*$   &     50    &   2.8    \\
S~22$^*$   &     49    &   6.0    \\
S~18   &     35    &   3.4    \\
R~50$^*$   &     81    &   1.5    \\
\noalign{\smallskip}
\hline
\end{tabular}
\renewcommand{\arraystretch}{1}
\begin{list}{}{}
\item[$^{\rm a}$] stellar radii (Zickgraf 1990)
\item[$^{\rm b}$] electron density values (Z89)
\item[$^{\rm c}$] values obtained in this work
\item[$^*$] edge-on stars
\end{list}
\end{table}

\subsection{Cylindrical geometry}

The  envelope with cylindrical geometry was modeled adopting a stellar radius of 
70~$R_{\sun}$, typical of a B[e]SG. The dimensions of the
cylindrical disk are the same used earlier (Sect. 3.1), namely, inner radius 
$R_{i}$ = 1~$R_{\star}$, outer radius $R_{o}$ = 300~$R_{\star}$ and thickness 
of 1~$R_{\star}$.

The electron density law used was:

\begin{equation}
n_e(r) = \frac{\rm{K}}{r^{n}} \,,
\end{equation}

\noindent where K is a constant and n is an exponent for constant ($n$=0) or variable
($n$=2) densities across the envelope. The constant K can be derived using the 
relation between the optical depth and the density law,

\begin{equation}
\tau = \int\limits^{R_o}_{R_i}\sigma_{\rm{T}} n_e(r)dr \,,
\end{equation}

\noindent with $\sigma_{\rm{T}}$ being the Thomson scattering cross section.

\subsubsection{Homogeneous density}

In Fig. 4 we show the intrinsic polarization as a function of mean electron density
(and equatorial optical depth) for a cylindrical disk with homogeneous density
for several inclination angles. From the figure, we can note that these models 
fail to fit the observational data as most data points, including the edge-on 
objects, lie well above the curve i=90$\degr$, which is
the maximum polarization predicted by the model. We conclude that homogeneous 
cylindrical envelopes are not a good representation for the MC B[e]SG envelopes.

\begin{figure}
\psfig{figure=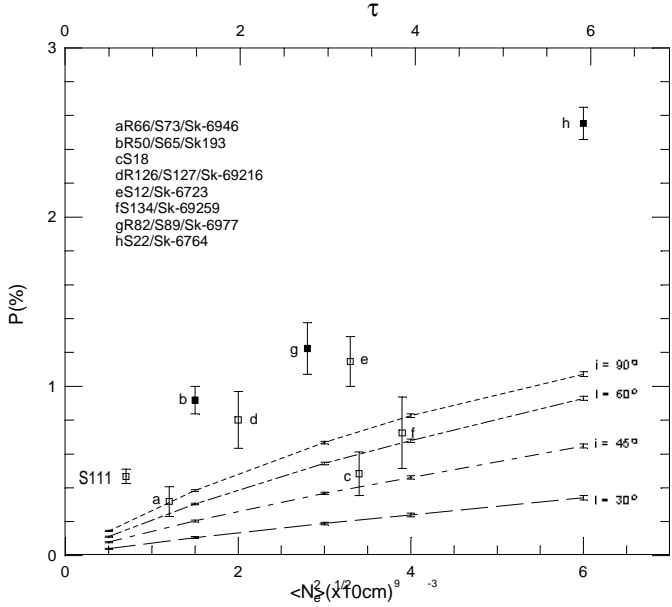,width=8.8cm,height=8cm}
\caption{Intrinsic polarization of MC B[e]SG as a function of envelope mean 
electron density $N_{e}$ (= ${<N_{e}^{2}> }^{1/2}$) and equatorial optical depth 
$\tau$. Lines represent models with cylindrical geometry and homogeneous electron 
density for several inclination angles. The points with error bars 
correspond to polarimetric data from this paper (S~111) and M92 
(other objects).}
\label{}
\end{figure}

\subsubsection{Density proportional to $r^{-2}$}

The model results when the density falls with $r^{-2}$ are presented 
in Fig. 5. In the figure,
most of the observational data points lie between the inclination angle curves
i=20\degr and i=45\degr. In particular, the edge-on objects fall far from the
i=90\degr curve. We conclude that, like the constant density models, the 
cylindrical $r^{-2}$ models do not fit well the observations either.

In order to investigate the dependence of the model results on the assumed 
stellar radius, we ran two additional cases, $R_{\star}$=30~$R_{\sun}$ and 
$R_{\star}$=140~$R_{\sun}$, which reasonably cover the range of MC B[e]SG 
radii (Table 3), for both constant and r$^{-2}$ density models. 
For the cases of constant density with $R_{\star}$=30 $R_{\sun}$, the model curves 
fall below all observed data points. The $R_{\star}$=140~$R_{\sun}$ model curves 
cover the $P-N_{e}$ data space better but still fail to adequately fit the edge-on 
cases. In addition, we know the edge-on objects do have in fact $R_{\star}$ between 
about 50 and 80~$R_{\sun}$ (Table 3). For the r$^{-2}$ density models, neither of 
these radius values fit the data points adequately.

\begin{figure} 
\psfig{figure=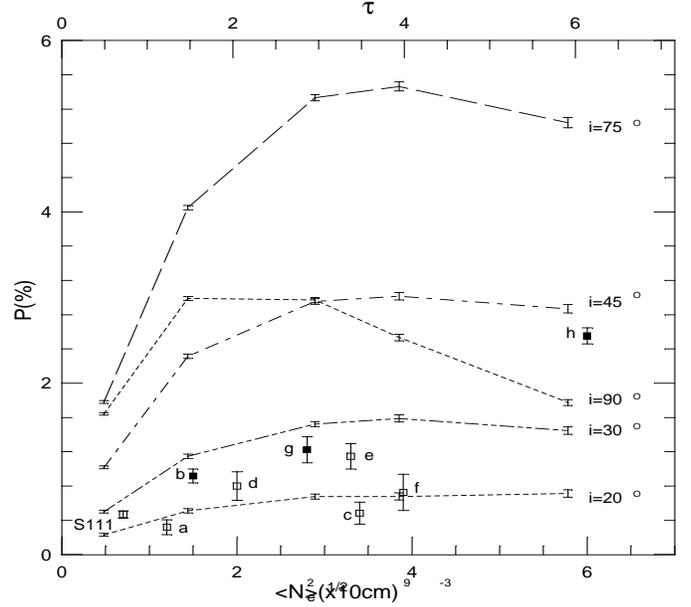,width=8.8cm,height=8cm}
\caption{Intrinsic polarization of MC B[e]SG as a function of mean electron
density square root $ \sqrt{<N_{e}^{2}> }$ and equatorial optical depth $\tau$. 
Data are as in Fig. 4. Lines represent models with cylindrical geometry and
electron density varying with $r^{-2}$ for several inclination angles.}
\label{}
\end{figure}

In summary, we conclude that the models using cylindrical geometry do not fit with 
the data for B[e]SG, at least not for the dimensions that have been proposed (Z85).

\subsection{Spherical geometry with density contrast}

To model the spherical
geometry with density contrast, we chose to use the parameterization of Waters et 
al. (1987), which has been also used by Wood et al. (1996). The
equatorial-to-polar density ratio may in fact render this model more 
realistic for the B[e]SG. 

The Waters et al. (1987) density law is

\begin{equation}
n_e(r) = n_0 \left(\frac{R_\star}{r}\right)^{n} (1 + A \sin^{m} \theta) \,,
\end{equation}

\noindent where $n_0$ is the density in the stellar surface, $\theta$ is the
polar angle. This parameterization consists of a spherically symmetric region
combined with an equatorial disk in which $(1 + A)$ determines the density
contrast, between the pole and equator, and $m$ is a measure of the opening 
angle of the disk. The aperture angle at which the density is half its maximum 
value (measured from the equator) is:

\begin{equation}
\theta_0 = \cos^{-1}(\frac{1}{2})^{1/m} .\
\end{equation}

We have chosen to model two cases, that of an envelope with a constant 
density ($n$=0), and that with density proportional to $r^{-2}$ ($n$=2).
In these cases, we used the same 
amount of mass as that of the scattering cylindrical disk. In doing that, the 
spherical envelope outer radius derived was 2850~$R_{\sun}$ 
($\sim$41~$R_{\star}$) . Again, the stellar 
radius used in the models was 70~$R_{\sun}$.

\subsubsection{Constant density}

In the case of homogeneous density, we see that in eq. 5, $n_e(r)$ only depends
on $\theta$, $m$ and $A$. We set the opening angle at $\theta$=5$\degr$, for which
$m$=182. Zickgraf (1989) estimated disk opening angles of 20$\degr$ to 40$\degr$. 
We have run models with this range of values but the resulting polarization leves were too
high (e.g., 7\% for 
$N_e$ = 6 $\times$ $10^9$ cm$^{-3}$). The value of density contrast used was 
$10^3$ (Z86), which implies in $A$=999 in eq. 5.

In Fig. 6, we present the results for envelopes with the geometry parameterized with 
eq. 5 and constant electron density. These models provide a good fit
to the observations, including the edge-on objects. We note that polarization 
level for i=90$\degr$ is larger than that obtained with cylindrical geometry 
(Fig. 4). This is due to 
the star, in the cylindrical geometry case, not being completely occulted by
the envelope. The star then contributes with direct, non-polarized flux, 
lowering the resulting polarization level.

\begin{figure}
\psfig{figure=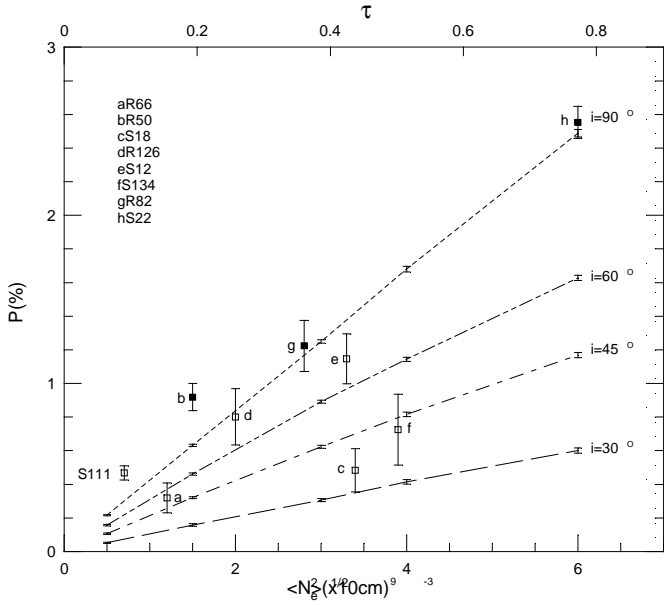,width=8.8cm,height=8cm}
\caption{Intrinsic polarization of MC B[e]SG as a function of electron
density $N_{e}$ (= ${<N_{e}^{2}> }^{1/2}$) and equatorial optical depth $\tau$. 
Data point are as in Fig. 4. Lines represent models with spherical geometry with 
density contrast and uniform electron density.}
\label{}
\end{figure}

\subsubsection{Density proportional to $r^{-2}$}

With the density now proportional to $r^{-2}$, we use the same parameter set 
as for constant density. The model results are presented in Fig. 7.
In the figure we can see that the model polarization obtained fall well below 
the expected level.

\begin{figure}
\psfig{figure=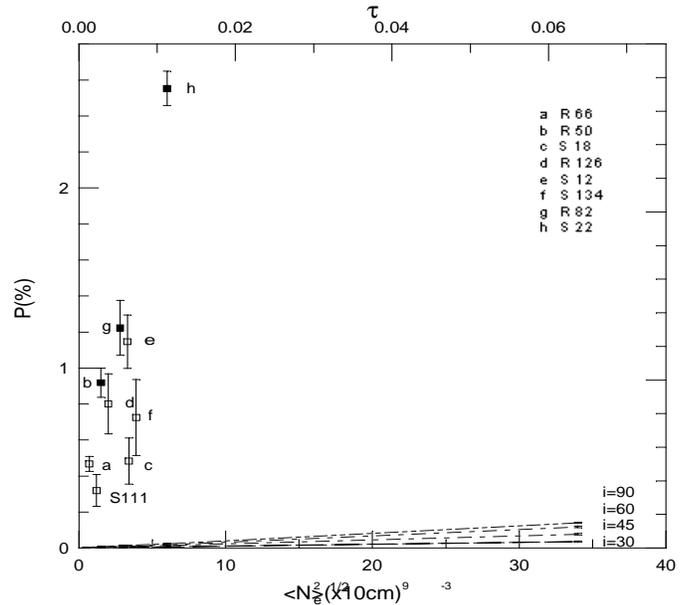,width=8.8cm,height=8cm}
\caption{Intrinsic polarization of MC B[e]SG as a function of mean electron
density square root $ \sqrt{<N_{e}^{2}> }$ and the equatorial optical depth 
$\tau$. Data are the same of Fig. 4. Here we considered a spherical geometry 
with density contrast, and a $r^{-2}$ electron density.}
\label{}
\end{figure}


\section{Conclusions}

We present optical polarization data for the LMC B[e]SG S~111.
The estimated intrinsic polarization is (0.467
$\pm$ 0.042)\% at 75$\fdg$3. This polarization is consistent with the presence
of a non-spherical circumstellar envelope around the object. 

After excluding AV~16/R~4, a B[e]SG now known to be a binary, and with 
the inclusion of S~111, we found using more recent IR calibration data that 
intrinsic polarization correlates better with the electron density
rather than the color excess $\Delta(K-L)$ in the envelopes of B[e]SG. This
suggests that electron scattering is the main mechanism that causes polarization
in the envelopes of these stars.

The spectroscopic data have been suggestive (Z85) of a two-component 
wind model for B[e]SG, i.e., a slow equatorial wind and a fast polar 
wind. We have hence considered two idealized models for such wind, one 
with a cylindrical geometry and another where the material is 
spherically distributed around the star but with a contrast enhancement 
along the equator. The results of our Monte Carlo modeling show that 
the spherical geometry with such density contrast fits the polarization 
observations very reasonably, being consistent with the two-component 
model for B[e]SG. In addition, the constant density case fits the data 
better than an $r^{-2}$ electron density law. This result agrees with 
the presence of a slow equatorial wind observed and modeled by 
Zickgraf et al. (1996b).

Questions still to be answered include details of the geometry of the 
equatorial disk, such as its opening angle, and the precise location of 
the scattering material within the disk. These will have to await 
spectropolarimetric observations of these objects. Changes (or the absence of
them) in polarization accross spectral features, e.g., emission lines,
would help us point out more precisely where the polarization arises in
the envelope (decreases across emission line, for instance, would
indicate that the polarization is produced closer to the star, etc.).
Other spectropolarimetric features, such as changes across the Balmer
discontinuities, might be helpful in assessing the envelope's opening
angle (e.g., Wood et al. 1997, for Be stars).

\begin{acknowledgements}
We thank the referee for suggestions which helped improve the paper.
RM acknowledges the CNPq fellowships. AMM and ACC acknowledge support 
from the S\~ao Paulo State funding agency FAPESP and CNPq.
\end{acknowledgements}

\end{document}